# Pebble games with algebraic rules[*]

Anuj Dawar and Bjarki Holm

University of Cambridge Computer Laboratory
`firstName.lastName@cl.cam.ac.uk`

November 10, 2018

**Abstract**

We define a general framework of *partition games* for formulating two-player pebble games over finite structures. We show that one particular such game, which we call the *invertible-map game*, yields a family of polynomial-time approximations of graph isomorphism that is strictly stronger than the well-known Weisfeiler-Lehman method. The general framework we introduce includes as special cases the pebble games for finite-variable logics with and without counting. It also includes a *matrix-equivalence game*, introduced here, which characterises equivalence in the finite-variable fragments of matrix-rank logic. We show that the equivalence defined by the invertible-map game is a refinement of the equivalence defined by each of these games for finite-variable logics.

## 1 Introduction

An important open problem in finite model theory is that of finding a logical characterisation of polynomial-time computability. That is to say, to find a logic in which a class of finite structures is expressible if and only if membership in the class is decidable in deterministic polynomial time (PTIME). The exact formulation of the problem (see [10]) requires additional effectivity conditions which need not concern us here. By a result proved independently by Immerman [13] and Vardi [18], it is known that inflationary fixed-point logic (IFP) expresses exactly the polynomial-time properties of *ordered* finite structures, but falls short of expressing the polynomial-time properties of *all* finite structures. It was at one time conjectured by Immerman that the extension of inflationary fixed-point logic with a mechanism for counting (IFPC) suffices for expressing all of PTIME. However, this turns out not to be the case and a counter-example was constructed by Cai, Fürer and Immerman [2]. Noting that this construction and various other examples of properties in PTIME that are not definable in IFPC can be reduced to testing the solvability of systems of linear equations, we introduced in [4] the extension of inflationary fixed-point logic with matrix-rank operators (IFPR). This logic strictly extends the expressive power of IFPC while still being contained in PTIME and it remains an open question whether there are polynomial-time properties that are not definable in IFPR.

---
[*]Research supported by EPSRC grant EP/H026835/1. An extended abstract of this paper will appear in the proceedings of ICALP 2012.

In this context, the study of ever more expressive logics has gone hand in hand with the development of tools for proving limitations on those logics. An important class of such tools are the so-called pebble games, which are variations and extensions of the Ehrenfeucht-Fraïssé game for first-order logic. In particular, the *k-pebble game* characterises the relation $\equiv_k^L$ of equivalence in first-order logic with $k$ variables. Since it can be shown that any formula of IFP is invariant under $\equiv_k^L$ for some $k$, this becomes useful in proving inexpressibility results for IFP. Similarly, inexpressibility results for IFPC are established by showing that a property is not invariant under $\equiv_k^C$ for any $k$, where $\equiv_k^C$ denotes the relation of equivalence in first-order logic with counting quantifiers and at most $k$ variables. The relation $\equiv_k^C$ has been characterised by two different pebble games: the Immerman-Lander game [14] and the *bijection game* of Hella [11].

In addition to providing a tool for the analysis of logics, these games also provide interesting approximations of the graph isomorphism relation. In particular, it can be shown that the equivalence relation $\equiv_{k+1}^C$ is exactly the relation decided by the $k$-dimensional Weisfeiler-Lehman method (see [2] for a description of the method and its relationship with $\equiv_k^C$). This is a family of polynomial-time algorithms which approach graph isomorphism in the limit by ever finer approximations. A key contribution of the Cai-Fürer-Immerman construction of a property in PTIME that is not definable in IFPC is to show that there is no fixed $k$ such that the $k$-dimensional Weisfeiler-Lehman algorithm decides graph isomorphism.

In a similar way, the logic of matrix-rank operators defined in [4] yields a family of equivalence relations $\equiv_{k,m,\Omega}^R$ which provide a stratification of graph isomorphism and which can be used to analyse definability in IFPR. Here, $\equiv_{k,m,\Omega}^R$ refers to equivalence in the logic that extends $k$-variable first-order logic with matrix-rank operators of arity at most $m$ for matrices over $\mathsf{GF}_p$ for any $p$ in the finite set of primes $\Omega$. One of our main contributions in this paper is a game that characterises this logical equivalence. This game, which we call the *matrix-equivalence game*, is difficult to use and it remains a challenge to deploy it to establish that there is a PTIME property not closed under $\equiv_{k,m,\Omega}^R$ for any $k$, $m$ and $\Omega$.

The matrix-equivalence game and the relations $\equiv_{k,m,\Omega}^R$ that it characterises suffer from another limitation as approximations of the graph isomorphism relation. Namely, it is not clear whether $\equiv_{k,m,\Omega}^R$ can be decided in polynomial time. Indeed, the natural algorithm that is obtained from the definition of the matrix-equivalence game runs in exponential time. This leads us to consider an alternative game that we call the *invertible-map game*. This game is obtained by replacing the algebraic matrix-equivalence condition with a condition of simultaneous similarity of tuples of matrices. As a result we obtain a family of equivalence relations $\approx_{m,\Omega}^k$ which refine $\equiv_{k,m,\Omega}^R$. Even though these relations are refinements of those obtained from the matrix-equivalence game, they seem easier to decide. Using a result of Chistov et al. [3] we are able to show that each of the relations $\approx_{m,\Omega}^k$ is decidable in polynomial time. Therefore, this gives us a family of *polynomial-time* algorithms which, like the Weisfeiler-Lehman method, approximates isomorphism in the limit. This family is strictly stronger than the Weisfeiler-Lehman method in the sense that it can also distinguish the Cai-Fürer-Immerman graphs at some fixed level.

The games we introduce in this paper are formulated as *partition games*. They are so called because the Duplicator is required at each move to give a suitable partition of the game board. This partition has to satisfy certain algebraic conditions which vary according to the game we are considering. It turns out that the games for $\equiv_k^L$ and $\equiv_k^C$ can be formulated as partition games, by replacing the algebraic rules of the matrix-equivalence game with weaker conditions that the partitions need to satisfy. This provides a general framework for exploring other games and, indeed, other equivalence relations on structures. So far, model-comparison games have been formulated for specific logics. Perhaps we can reverse this and extract suitable logics from well-behaved games? One such challenge is to formulate a logic that corresponds to the invertible map game that we define here.

## 2 Preliminaries

We assume that all structures are finite and that all vocabularies are finite and relational. Throughout, we commonly write $\tau$ to denote a vocabulary. We write $U(\mathbf{A})$ for the universe of a structure $\mathbf{A}$ and write $\|\mathbf{A}\|$ for the cardinality of $U(\mathbf{A})$. We denote the class of all finite $\tau$-structures with fixed tuples of $r \in \mathbb{N}$ parameters by $\mathrm{fin}[\tau; r] := \{(\mathbf{A}, \vec{a}) \mid \mathbf{A} \in \mathrm{fin}[\tau],\ \vec{a} \in U(\mathbf{A})^r\}$. We denote tuples $(v_1, \ldots, v_k)$ by $\vec{v}$ and their length by $\|\vec{v}\|$. If $\vec{v}$ is a $k$-tuple of elements from a set $X$, $i \in [k]$ and $w \in X$, then we write $\vec{v}\frac{w}{i}$ for the tuple obtained from $\vec{v}$ by replacing the $i$-th component with $w$; that is, $\vec{v}\frac{w}{i} = (v_1, \ldots, v_{i-1}, w, v_{i+1}, \ldots, v_k)$. If $m \leq k$, $\vec{i} = (i_1, \ldots, i_m) \in [k]^m$ is a tuple of distinct integers (an 'index pattern') and $\vec{w}$ is an $m$-tuple of elements from $X$, then we write $\vec{v}\frac{\vec{w}}{\vec{i}} := \vec{v}\frac{w_1}{i_1} \cdots \frac{w_m}{i_m}$.

### 2.1 Matrices indexed by unordered sets

If $F$ is a field and $I$, $J$ are finite and non-empty sets then an $I \times J$ *matrix* over $F$ is a function $M : I \times J \to F$. If $\|I\| = \|J\|$ then we say that $M$ is *invertible* if there is a $J \times I$ matrix $N$ such that the product $MN$ is the $I \times I$ identity matrix; that is, if $MN(x, y) = 1$ if $x = y$ and $MN(x, y) = 0$ otherwise [1]. In this case we refer to $N$ as the inverse of $M$, denoted by $M^{-1}$. A matrix whose rows and columns are indexed by the same set is said to be *square*. Recall that is $M$ is a square $I \times I$ matrix, $N$ is a square $J \times J$ matrix and $\|I\| = \|J\|$, then we say that $M$ and $N$ are *equivalent* if there is an invertible $J \times I$ matrix $P$ and an invertible $I \times J$ matrix $Q$ such that $PMQ = N$. The two matrices $M$ and $N$ are said to be *similar* if there is an invertible $J \times I$ matrix $S$ such that $SMS^{-1} = N$. The transformation $M \mapsto SMS^{-1}$ is called a *similarity transformation* by the similarity matrix $S$.

In this paper we focus on square $\{0, 1\}$-matrices whose rows and columns are indexed by tuples of elements from some finite and non-empty base set $A$. More specifically, if $B \subseteq A^{2m}$ for some $m \geq 1$, then we write $\chi_B$ for the characteristic function of $B$, seen as a $\{0, 1\}$-matrix indexed by $A^m \times A^m$. That is, $\chi_B$ is defined by $(\vec{a}, \vec{b}) \mapsto 1$ if $(\vec{a}, \vec{b}) \in B$ and $(\vec{a}, \vec{b}) \mapsto 0$ otherwise. We refer to $\chi_B(\vec{a}, \vec{b})$ as the *characteristic matrix of B*; the underlying field and the exponent $m$ are usually clear from the context.

---

[1] Equivalently, it can be seen that $M$ is invertible if there is an $J \times I$ matrix $N$ such that $NM$ is the $J \times J$ identity matrix.

We also consider matrices expressed as a linear combination of characteristic matrices. Let $\mathbf{P} \subseteq \wp(A^{2m})$ be a non-empty collection of subsets of $A^{2m}$ and let $\gamma : \mathbf{P} \to F$ be a function. Then we write $M_\gamma^{\mathbf{P}}$ to denote the $A^m \times A^m$ matrix over $F$ defined by $M_\gamma^{\mathbf{P}} := \sum_{P \in \mathbf{P}} \gamma(P) \cdot \chi_P$. Typically, $\mathbf{P}$ will be a *partition* of $A^{2m}$; that is, a collection of non-empty and mutually disjoint subsets of $A^{2m}$ (called *blocks*) whose union is all of $A^{2m}$.

## 2.2 Finite-variable logics.

We write $L^k$ to denote the fragment of first-order logic using only the variables $x_1, \ldots, x_k$ and we write $C^k$ for the extension of $L^k$ with rules for defining counting formulae of the kind $\exists^{\geq i} x \,.\, \varphi(x)$, for $i > 0$ (for further details, see [7, 16]). For $(\mathbf{A}, \vec{a})$ and $(\mathbf{B}, \vec{b})$ in fin$[\tau; r]$, we write $(\mathbf{A}, \vec{a}) \equiv_k^L (\mathbf{B}, \vec{b})$ to indicate that for any $L^k$-formula $\varphi$, it holds that $(\mathbf{A}, \vec{a}) \models \varphi$ if and only if $(\mathbf{B}, \vec{b}) \models \varphi$; the relation $\equiv_k^C$ is defined similarly for $C^k$.

For each integer $i > 0$ and prime $p$, we define a quantifier $\mathrm{rk}_p^{\geq i}$ which binds exactly $2m$ variables. If $\varphi_1, \ldots, \varphi_{p-1}$ are formulae, $\vec{x}$ and $\vec{y}$ are $m$-tuples of pairwise distinct variables, and $\mathbf{A}$ a structure, then we let

$$\mathbf{A} \models \mathrm{rk}_p^{\geq i}(\vec{x}, \vec{y}) \,.\, (\varphi_1, \ldots, \varphi_{p-1})$$

if and only if the rank of the square $U(\mathbf{A})^m \times U(\mathbf{A})^m$ matrix $\sum_{j=1}^{p-1} j \cdot \chi_{\varphi_j^{\mathbf{A}}}$ (mod $p$) is at least $i$ over $\mathsf{GF}_p$, the finite field with $p$ elements. If $\Omega$ is a finite and non-empty set of primes, then we write $R_{m;\Omega}^k$ to denote the logic built up in the same way as $k$-variable first-order logic, except that we have rules for constructing formulae with $2m$-ary rank quantifiers over $\mathsf{GF}_p$ ($p \in \Omega$) instead of the rules for first-order existential and universal quantifiers. Every formula in $L^k$ or $C^k$ is equivalent to one of $R_{\{p\};2}^{k+1}$ (where $p$ is any prime), for we can simulate existential, universal and unary counting quantifiers by expressing the rank of diagonal matrices (see [4, 12, 15] for details). We write $(\mathbf{A}, \vec{a}) \equiv_{k,m,\Omega}^R (\mathbf{B}, \vec{b})$ to indicate that $\mathbf{A}$ and $\mathbf{B}$ agree on all $R_{m;\Omega}^k$-formulae under the assignments $\vec{a}$ and $\vec{b}$, respectively. It can be shown that any formula of IFPR is *invariant* under $\equiv_{k,m,\Omega}^R$ for some $k$, $m$ and $\Omega$ [4, 12, 15]. This means that for any IFPR-formula $\varphi(\vec{x})$ there are $k$, $m$ and $\Omega$, such that if $(\mathbf{A}, \vec{a})$ and $(\mathbf{B}, \vec{b})$ are structures such that $(\mathbf{A}, \vec{a}) \models \varphi$ and $(\mathbf{A}, \vec{a}) \equiv_{k,m,\Omega}^R (\mathbf{B}, \vec{b})$ then also $(\mathbf{B}, \vec{b}) \models \varphi$.

*Remark.* An alternative way to define finite-variable rank logic is to extend $L^k$ with all rank quantifiers of arity *up to* some fixed bound, including quantifiers for non-square matrices as well as the usual first-order quantifiers $\exists$ and $\forall$. This would generally give us tighter bounds on the number of variables required to express properties using rank quantifiers. However, our interest is not so much in these finite-variable logics themselves, but rather in using them as tools for showing inexpressibility in fixed-point logic with rank operators, which means showing non-definability in $R_{m;\Omega}^k$ for *all* values of $k$, $m$ and $\Omega$. Allowing different forms of quantifiers in $R_{m;\Omega}^k$ would therefore only serve to complicate the games we develop in this paper, without providing any new insight.

## 2.3 Pebble games

Definability in $L^k$ is elegantly characterised in terms of two-player games based on a game style originally developed by Ehrenfeucht and Fraïssé [9, 8]. These games were essentially given by Barwise [1] though versions were also independently presented by Immerman [13] and Poizat [17]. The game board of the *k-pebble game* consists of two structures $\mathbf{A}$ and $\mathbf{B}$ over the same vocabulary and $k$ pebbles for each of the two structures, labelled $1, \ldots, k$. The game has two players, Spoiler and Duplicator. At each round of the game, the following takes place.

1. Spoiler picks up a pebble in one of the structures (either an unused pebble or one that is already on the board) and places it on an element of the corresponding structure. For instance he[2] might take the pebble labelled by $i$ in $\mathbf{B}$ and place it on an element of $\mathbf{B}$.

2. Duplicator must respond by placing the matching pebble in the opposite structure. In the above example, she must place the pebble labelled by $i$ on an element of $\mathbf{A}$.

Assume at the end of the round that $r$ pebbles have been placed and let $\{(a_i, b_i) \mid 1 \leq i \leq r\} \subseteq U(\mathbf{A}) \times U(\mathbf{B})$ denote the $r$ pairs of pebbled elements, such that for each $i$ the label of the pebble on element $a_i$ is the same as the label of the pebble on element $b_i$. If the partial map $f : U(\mathbf{A}) \to U(\mathbf{B})$ given by

$$f := \{(a_i, b_i) \mid 1 \leq i \leq r\} \cup \{(c^{\mathbf{A}}, c^{\mathbf{B}}) \mid c \in \tau \text{ a constant}\}$$

is not a partial isomorphism, then Spoiler has won the game; otherwise it can continue for another round. We say that Duplicator has a winning strategy in the $k$-pebble game if she can play the game forever, maintaining a partial isomorphism at the end of each round. We also consider the situation where the game starts with some of the pebbles initially placed on the game board. Formally, we refer to a placement of pebbles over one of the structures as a *position*. If $\vec{a}$ and $\vec{b}$ are $r$-tuples of elements from $U(\mathbf{A})$ and $U(\mathbf{B})$ respectively, $r \leq k$, then the game starting with positions $(\mathbf{A}, \vec{a})$ and $(\mathbf{B}, \vec{b})$ is played as above, except that pebbles $1, \ldots, r$ in $\mathbf{A}$ are initially placed on the elements $a_1, \ldots, a_r$ of $\vec{a}$ and pebbles $1, \ldots, r$ in $\mathbf{B}$ are initially placed on the elements $b_1, \ldots, b_r$ of $\vec{b}$.

The result that links this game with definability in $L^k$ says that Duplicator has a winning strategy in the $k$-pebble game starting with positions $(\mathbf{A}, \vec{a})$ and $(\mathbf{B}, \vec{b})$ if and only if $(\mathbf{A}, \vec{a}) \equiv_k^L (\mathbf{B}, \vec{b})$. This correspondence gives us the a purely combinatorial method for proving inexpressibility results for $k$-variable logic in general and IFP in particular, since it can be shown that any formula of IFP is invariant under $\equiv_k^L$ for some $k$. This method can be formalised as follows:

> To show that a property $P$ of finite structures is not definable in IFP, it suffices to show that for each $k \in \mathbb{N}$ there is a pair of structures $(\mathbf{A}_k, \mathbf{B}_k)$ for which it holds that: (*i*) $\mathbf{A}_k$ has property $P$ but $\mathbf{B}_k$ does not; and (*ii*) Duplicator has a winning strategy in the $k$-pebble game on $\mathbf{A}_k$ and $\mathbf{B}_k$.

---
[2] By convention, Spoiler is male and Duplicator female.

Immerman and Lander [14] and Hella [11] later introduced separate versions of the $k$-pebble game for analysing the expressiveness of $C^k$ over finite models. Both of these games can be used to establish lower bounds for IFPC over finite structures since it can be shown that any formula of IFPC is invariant under $\equiv_k^C$ for some $k$. Here we focus our attention on the game given by Hella, which we refer to as the *$k$-pebble bijection game*. As before, the game is played by Spoiler and Duplicator (each with $k$ pebbles) on structures $\mathbf{A}$ and $\mathbf{B}$. If $\|\mathbf{A}\| \neq \|\mathbf{B}\|$, then Spoiler wins the game immediately. Otherwise, each round of the game proceeds as follows:

1. Spoiler picks up a pebble from $\mathbf{A}$ and the matching pebble from $\mathbf{B}$.

2. Duplicator has to respond by choosing a bijection $h : U(\mathbf{A}) \to U(\mathbf{B})$.

3. Spoiler then places the pebble chosen from $\mathbf{A}$ on some element $a \in U(\mathbf{A})$ and places the matching pebble from $\mathbf{B}$ on $h(a)$.

This completes one round in the game. If, after this round, the partial map from $\mathbf{A}$ to $\mathbf{B}$ defined by the pebbled positions (plus constants) is not a partial isomorphism, then Spoiler has won the game. Otherwise it can continue for another round. This game characterises definability in $C^k$ in the sense that Duplicator has a winning strategy in the $k$-pebble bijection game starting with positions $(\mathbf{A}, \vec{a})$ and $(\mathbf{B}, \vec{b})$ if and only if $(\mathbf{A}, \vec{a}) \equiv_k^C (\mathbf{B}, \vec{b})$.

## 2.4 Class extensions and extension matrices

We frequently consider relations that arise by extending a fixed tuple of elements in a structure according to some criteria. For example, consider a formula $\varphi$ and let $\vec{a}$ be an assignment of values to the free variables of $\varphi$ over a structure $\mathbf{A}$. Then the set of all pairs $(c, d)$ from $\mathbf{A}$ which, when used to replace the first two elements of $\vec{a}$ to give a satisfying assignment to $\varphi$, can be seen as a binary "extension" of $\vec{a}$ in $\mathbf{A}$, defined by the formula $\varphi$. Moreover, this relation can be viewed as a $\{0, 1\}$-matrix over $\mathbf{A}$ in the usual way, which gives us a way to associate a pair $(\mathbf{A}, \vec{a})$ with a family of matrices over $\mathbf{A}$.

More formally, consider a class $\alpha \subseteq \text{fin}[\tau; k]$ and let $\vec{i} = (i_1, \ldots, i_n) \in [k]^n$ be a tuple of distinct integers, $n \leq k$. Then we write $\text{ext}_{\vec{i}}^{\alpha}$ to denote the functor on $\text{fin}[\tau; k]$ defined by $\text{ext}_{\vec{i}}^{\alpha}(\mathbf{A}, \vec{a}) := \{\vec{b} \in U(\mathbf{A})^n \mid (\mathbf{A}, \vec{a}\frac{\vec{b}}{\vec{i}}) \in \alpha\}$. We refer to $\text{ext}_{\vec{i}}^{\alpha}(\mathbf{A}, \vec{a})$ as the *$\vec{i}$-extension of $(\mathbf{A}, \vec{a})$ into $\alpha$*. Abusing notation, if $\varphi$ is a formula whose free variables are all amongst $\vec{x} = (x_1, \ldots, x_k)$, then we let $\text{ext}_{\vec{i}}^{\varphi} := \text{ext}_{\vec{i}}^{\alpha_\varphi}$, where $\alpha_\varphi := \{(\mathbf{A}, \vec{a}) \in \text{fin}[\tau; k] \mid (\mathbf{A}, \vec{a}) \models \varphi\}$. That is,

$$\text{ext}_{\vec{i}}^{\varphi} = \{\vec{b} \in U(\mathbf{A})^n \mid \mathbf{A} \models \varphi[\vec{a}\frac{\vec{b}}{\vec{i}}]\} \subseteq U(\mathbf{A})^n.$$

If $n = 2m$, then we write $\text{extmat}_{\vec{i}}^{\alpha}(\mathbf{A}, \vec{a})$ and $\text{extmat}_{\vec{i}}^{\varphi}(\mathbf{A}, \vec{a})$ to denote the $U(\mathbf{A})^m \times U(\mathbf{A})^m$ characteristic matrices of $\text{ext}_{\vec{i}}^{\alpha}(\mathbf{A}, \vec{a})$ and $\text{ext}_{\vec{i}}^{\varphi}(\mathbf{A}, \vec{a})$, respectively. We refer to such matrices as *extension matrices*.

## 3 A game characterisation of rank logics

In order to analyse the expressive power of rank logics over finite structures, it is important to develop methods for proving non-definability. In this context, the restriction to finite structures means that many of the classical tools of model theory, such as the compactness theorem, are not available. Instead, we consider extensions of pebble games—variations of Ehrenfeucht-Fraïssé games for first-order logic—which have assumed a central role in the study of both finite-variable and fixed-point logics. In this section we give a pebble-game characterisation of finite-variable logic with quantifiers for matrix rank. This gives us a combinatorial method for proving lower bounds (inexpressibility results) for fixed-point logic with rank operators, thereby settling one of the open problems presented in [4].

To give the intuition behind this game, we first describe a simple "partition game" that is a based on the same game protocol. The partition game is played by two players, Spoiler and Duplicator, on a pair of relational structures $\mathbf{A}$ and $\mathbf{B}$, each with $k$ pebbles labelled $1, \ldots, k$. At each round of the game, Spoiler removes a pebble from $\mathbf{A}$ and the corresponding pebble from $\mathbf{B}$. Unlike the classical pebble game, Duplicator is not allowed to move any pebbles herself (by convention, Spoiler is male and Duplicator female). However, in response to the challenge of the Spoiler, she is allowed to divide the game board into disjoint regions in order to restrict the possible moves that Spoiler is subsequently allowed to make. More specifically, in response to Spoiler's challenge, Duplicator partitions each of $U(\mathbf{A})$ and $U(\mathbf{B})$ into the same number of disjoint regions and gives a matching between the regions in $U(\mathbf{A})$ and the regions in $U(\mathbf{B})$. Intuitively, Duplicator's strategy will be to gather in each region all those elements that lead to game positions that are sufficiently alike. In turn, Spoiler is allowed to place each of the chosen pebbles on some element of the corresponding structure, with the *restriction* that the two newly pebbled elements have to be within matching regions. That completes a round of the game. Compared with the standard pebble game, it may seem that the partition game is biased against the Duplicator, since she is not allowed to place her own pebbles after seeing where Spoiler places his. However, it can be shown that the two games are actually equivalent over finite structures.

The idea of dividing the game board into disjoint regions leads to a very generic template for designing new pebble games. For instance, if we adapt the rules so that any two matching regions need to have the same cardinality, then we get a game equivalent to the bijection game. The "matrix-equivalence game" we describe next is obtained by putting additional linear-algebraic constraints on the matching game regions.

## 3.1 Matrix-equivalence game

Let $k$ and $m$ be positive integers with $2m \leq k$ and let $\Omega$ be a finite and non-empty set of primes. The game board of the $k$-pebble $m$-ary *matrix-equivalence game* over $\Omega$ (or $(k, m, \Omega)$-matrix-equivalence game for short) consists of two structures $\mathbf{A}$ and $\mathbf{B}$ of the same vocabulary, each with $k$ pebbles labelled $1, \ldots, k$. The first $r \leq k$ pebbles of $\mathbf{A}$ may be initially placed on the elements of an $r$-tuple $\vec{a}$ of elements in $\mathbf{A}$ and the corresponding $r$ pebbles in $\mathbf{B}$ on an $r$-tuple $\vec{b}$ of elements in $\mathbf{B}$. If $\|\mathbf{A}\| \neq \|\mathbf{B}\|$ or the mapping defined by the initial pebble positions is not a partial isomorphism then Spoiler wins the game immediately. Otherwise, each round of the game proceeds as follows.

1. Spoiler chooses a prime $p \in \Omega$ and picks up $2m$ pebbles in some order from **A** and the $2m$ corresponding pebbles in the same order from **B**.

2. Duplicator has to respond by choosing

    - a partition **P** of $U(\mathbf{A})^m \times U(\mathbf{A})^m$,
    - a partition **Q** of $U(\mathbf{B})^m \times U(\mathbf{B})^m$, with $\|\mathbf{P}\| = \|\mathbf{Q}\|$, and
    - a bijection $f : \mathbf{P} \to \mathbf{Q}$,

    for which it holds that for all labellings $\gamma : \mathbf{P} \to \mathsf{GF}_p$,

    $$\mathrm{rk}(M^{\mathbf{P}}_\gamma) = \mathrm{rk}(M^{\mathbf{Q}}_{\gamma \circ f^{-1}}). \tag{r.c.}$$

    Here the composite map $\gamma \circ f^{-1} : \mathbf{Q} \to \mathsf{GF}_p$ is seen as a labelling of **Q**.

3. Spoiler next picks a block $P \in \mathbf{P}$ and places the $2m$ chosen pebbles from **A** on the elements of some tuple in $P$ (in the order they were chosen earlier) and the corresponding $2m$ pebbles from **B** on the elements of some tuple in $f(P)$ (in the same order).

This completes one round in the game. If, after this exchange, the partial map from **A** to **B** defined by the pebbled positions (in addition to constants) is not a partial isomorphism, or if Duplicator is unable to produce the required partitions, then Spoiler wins the game; otherwise it can continue for another round. We say that Duplicator has a *winning strategy* in the game if she can continue playing forever, maintaining a partial isomorphism at the end of each round.

*Remark.* Observe that the game condition "$\mathrm{rk}(M^{\mathbf{P}}_\gamma) = \mathrm{rk}(M^{\mathbf{Q}}_{\gamma \circ f^{-1}})$" is the same as saying that the two matrices should be *equivalent*, since rank is a complete invariant for matrix equivalence. This explains the name of the game.

Our main result here is the following theorem, which relates definability in finite-variable rank logic with a winning strategy for Duplicator in the matrix-equivalence game. A proof of the theorem is given in Section 3.2.

**Theorem 3.1.** *Duplicator has a winning strategy in the $(k, m, \Omega)$-matrix-equivalence game on $(\mathbf{A}, \vec{a})$ and $(\mathbf{B}, \vec{b})$ if and only if $(\mathbf{A}, \vec{a}) \equiv^R_{k,m,\Omega} (\mathbf{B}, \vec{b})$.*

By considering initial positions $(\mathbf{A}, \vec{a})$ and $(\mathbf{B}, \vec{b})$ where $\vec{a}$ and $\vec{b}$ are empty tuples (that is, when all pebbles are off the board at the start of the game), we get the following result.

**Corollary 3.2.** *Duplicator has a winning strategy in the $(k, m, \Omega)$-matrix-equivalence game on **A** and **B** if and only if $\mathbf{A} \equiv^R_{k,m,\Omega} \mathbf{B}$.*

The next lemma shows that for all $m$ and for all finite sets of primes $\Omega$, the equivalence on $\mathrm{fin}[\tau; k]$ given by the matrix-equivalence game refines that given by the bijection game. The proof follows from Theorem 3.1 by observing that $(i)$ the relation $\equiv^C_k$ is characterised by the $k$-pebble bijection game [11] and $(ii)$ we can simulate counting quantifiers by applying rank quantifiers to diagonal matrices (this was discussed further in Section 2). The increase in the number of pebbles needed from $k$ to $k + 2m - 1$ is due to this simulation of counting quantifiers by rank quantifiers: to replace a counting quantifier binding a single variable we need to introduce a rank quantifier binding $2m$ variables, so in general $2m - 1$ additional variables are needed for the translation.

**Lemma 3.3.** *If Duplicator has a winning strategy in the $(k + 2m - 1, m, \Omega)$-matrix-equivalence game on $(\mathbf{A}, \vec{a})$ and $(\mathbf{B}, \vec{b})$ for some $\Omega$ and $m \in \mathbb{N}$, then she has a winning strategy in the $k$-pebble bijection game starting on $(\mathbf{A}, \vec{a})$ and $(\mathbf{B}, \vec{b})$.*

From the viewpoint of finite model theory, the interest in studying the finite-variable logics $R_{m;\Omega}^k$ is mainly to analyse the expressive power of fixed-point logics with operators for matrix rank. In this context, the correspondence between $\equiv_{k,m,\Omega}^R$ and the matrix-equivalence game gives us a game-based method for proving non-definability of queries in IFPR. Compared with the game methods for IFP and IFPC that we discussed before, this proof method is however complicated by the fact that we need to consider two additional parameters—the set of primes $\Omega$ and the quantifier arity $m$—in addition to the number of variables $k$ employed in the game:

> To show that a property $P$ of finite structures is not definable in IFPR, it suffices to show for each $k, m \in \mathbb{N}$, with $2m \leq k$, and each finite set of primes $\Omega$ that there is a pair of structures $(\mathbf{A}_{k,m,\Omega}, \mathbf{B}_{k,m,\Omega})$ for which it holds that
>
> 1. $\mathbf{A}_{k,m,\Omega}$ has property $P$ but $\mathbf{B}_{k,m,\Omega}$ does not; and
> 2. Duplicator has a winning strategy in the $(k, m, \Omega)$-matrix-equivalence game on $\mathbf{A}_{k,m,\Omega}$ and $\mathbf{B}_{k,m,\Omega}$.

Finally, note that it requires much more effort to describe a winning strategy for Duplicator in the matrix-equivalence game compared with the pebble games we saw earlier. Based only on the pebbles chosen by Spoiler at the beginning of a round, Duplicator has to partition the two sides of the game board in a way that both satisfies the rank condition (**r.c.**) and which gives a satisfying response to *any* placement of pebbles by Spoiler in the subsequent move. Note in particular that once Duplicator has specified the partitions, she has no further input for the remainder of that game round.

## 3.2 Proof of the game characterisation

To simplify our notation, for the proof of Theorem 3.1 we consider only positions $(\mathbf{A}, \vec{a})$ and $(\mathbf{B}, \vec{b})$ with $\|\vec{a}\| = \|\vec{b}\| = k$; that is, positions where all the pebbles are initially placed on the board. The argument for the case when the tuples $\vec{a}$ and $\vec{b}$ have length $r < k$ is exactly the same, except that one has to distinguish at every turn between game moves made during the first $k$ rounds and game moves in the subsequent rounds[3]. This has the effect of making the proof non-uniform, without actually providing any new insight.

---

[3]Note that it is possible to obtain a proof for $\|\vec{a}\| = \|\vec{b}\| = r \in [k-1]$ as a direct corollary of the situation when $\|\vec{a}\| = \|\vec{b}\| = k$. In this case, given $r$-tuples $\vec{a}$ and $\vec{b}$, one would consider the game with pebble positions $\vec{a}'$ and $\vec{b}'$, where the $k$-tuple $\vec{a}'$ is obtained from $\vec{a}$ by adding $k-r$ copies of $a_1$ at the end of the tuple (simulating the case when $k-r+1$ pebbles are placed on element $a_1$) and similarly for $\vec{b}'$. Alternatively, one could consider a game board where the structures $\mathbf{A}$ and $\mathbf{B}$ are augmented with new vertices $\star_A$ and $\star_B$, respectively, disjoint

In order to give the proof, we first need to establish some technical results and new notation. Consider a tuple $\Phi = (\varphi_1, \ldots, \varphi_{p-1})$ of formulae in vocabulary $\tau$ and suppose that all the formulae occurring in $\Phi$ have free variables amongst the elements of the $k$-tuple $\vec{x}$. Let $\vec{i} = (i_1, \ldots, i_{2m})$ be a tuple of distinct integers, with $2m \leq k$. Then we write $\mathrm{fmat}_{\vec{x},\vec{i}}(\Phi, \mathbf{A}, \vec{a})_p$ to denote the "formula matrix" over $\mathsf{GF}_p$ defined by

$$\mathrm{fmat}_{\vec{x},\vec{i}}(\Phi, \mathbf{A}, \vec{a})_p := \sum_{i=1}^{p-1} i \cdot \mathrm{extmat}_{\vec{i}}^{\varphi_i}(\mathbf{A}, \vec{a}) \pmod{p}.$$

That is, $\mathrm{fmat}_{\vec{x},\vec{i}}(\Phi, \mathbf{A}, \vec{a})_p$ is a linear combination of the $U(\mathbf{A})^m \times U(\mathbf{A})^m$ extension matrices of the formulae $\varphi_i$, with scalar coefficients defined by the position of each formula in the tuple $\Phi$.

**Lemma 3.4.** *Suppose $(\mathbf{A}, \vec{a}) \equiv_{k,m,\Omega}^R (\mathbf{B}, \vec{b})$, with $\vec{a}$ and $\vec{b}$ $k$-tuples of elements. Let $\vec{x}$ be a $k$-tuple of variables and suppose $2m \leq k$ and $p \in \Omega$. Then for all $\varphi_1, \ldots, \varphi_{p-1} \in R_{m;\Omega}^k$, with $\mathrm{free}(\varphi_i) \subseteq \vec{x}$, and all tuples $\vec{i} \in [k]^{2m}$ of distinct integers, it holds that*

$$\mathrm{rk}(\mathrm{fmat}_{\vec{x},\vec{i}}(\Phi, \mathbf{A}, \vec{a})_p) = \mathrm{rk}(\mathrm{fmat}_{\vec{x},\vec{i}}(\Phi, \mathbf{B}, \vec{b})_p),$$

*where the matrix rank is taken over $\mathsf{GF}_p$ and $\Phi := (\varphi_1, \ldots, \varphi_{p-1})$.*

*Proof.* Let $(\mathbf{A}, \vec{a}) \in \mathrm{fin}[\tau; k]$. Then for all tuples $\Phi = (\varphi_1, \ldots, \varphi_{p-1})$ of $R_{m;\Omega}^k$-formulae, with $\mathrm{free}(\varphi_i) \subseteq \vec{x}$, and all $\vec{i} \in [k]^{2m}$, the formula

$$\mathrm{rk}_p^{=l}((x_{i_1}, \ldots, x_{i_m}), (x_{i_{m+1}}, \ldots, x_{i_{2m}})) \cdot (\varphi_1, \ldots, \varphi_{p-1})$$

is in $\mathrm{tp}(R_{m;\Omega}^k; \mathbf{A}, \vec{a})$ exactly for the number $l := \mathrm{rk}(\mathrm{fmat}_{\vec{x},\vec{i}}(\Phi, \mathbf{A}, \vec{a})_p)$. The statement of the lemma now follows by considering that $\mathrm{tp}(R_{m;\Omega}^k; \mathbf{A}, \vec{a}) = \mathrm{tp}(R_{m;\Omega}^k; \mathbf{B}, \vec{b})$. □

We are now ready to prove Theorem 3.1. The proof is given by two separate lemmas, one for each implication.

**Lemma 3.5.** *If $(\mathbf{A}, \vec{a}) \not\equiv_{k,m,\Omega}^R (\mathbf{B}, \vec{b})$ then Spoiler has a winning strategy in the $(k, m, \Omega)$-matrix-equivalence game starting with positions $(\mathbf{A}, \vec{a})$ and $(\mathbf{B}, \vec{b})$.*

In the proof of this lemma, we show that if $(\mathbf{A}, \vec{a}) \not\equiv_{k,m,\Omega}^R (\mathbf{B}, \vec{b})$ then Spoiler has a strategy to force the game, in a finite number of rounds, into positions that are not partially isomorphic. Spoiler's strategy is obtained by structural induction on some formula $\varphi \in R_{m;\Omega}^k$ on which the two game positions disagree; this argument broadly resembles similar proofs for the standard pebble games (see e.g. [16]). The main difficulty of the proof is to show that if Duplicator produces partitions $\mathbf{P}$ and $\mathbf{Q}$, then Spoiler can always find a block in one of the partitions that contains both tuples that satisfy $\varphi$ and tuples that satisfy $\neg \varphi$. Once he has identified such a block, Spoiler can place his pebbles in a way that ensures that the resulting game positions disagree on a formula of quantifier rank less than $\varphi$. This gives him a strategy to win the game in a finite number of moves.

---

from the rest of the structure. Here the idea is that a pebble placed on these special elements is to be treated as being off-the-board. This latter approach has the benefit of working for all $r \leq k$, including $r = 0$, without changing the proof in any other way. See for example Ebbinghaus and Flum [7] for an application of this idea.

*Proof.* If $(\mathbf{A}, \vec{a}) \not\equiv^R_{k,m,p} (\mathbf{B}, \vec{b})$ then there is a formula $\varphi(\vec{x}) \in R^k_{m;p}$ of quantifier rank $q \in \mathbb{N}$ such that $\mathbf{A} \models \varphi[\vec{a}]$ but $\mathbf{B} \models \neg\varphi[\vec{b}]$. If $q = 0$ then the mapping $\mathbf{A} \to \mathbf{B}$ defined by the pebbled elements $\vec{a} \mapsto \vec{b}$ is not a partial isomorphism and Spoiler has won the game. For the inductive step, suppose that $q > 0$. We show that Spoiler can in one round force the game into positions $(\mathbf{A}, \vec{a}')$ and $(\mathbf{B}, \vec{b}')$ where $(\mathbf{A}, \vec{a}')$ and $(\mathbf{B}, \vec{b}')$ can be distinguished by a formula of quantifier rank $q' < q$. This gives Spoiler a strategy to win the game in a finite number of steps, as claimed. To establish the claim, we can assume without loss of generality that $\varphi$ is of the form

$$\mathrm{rk}_p^{=l}((x_{i_1}, \ldots, x_{i_m}), (x_{i_{m+1}}, \ldots, x_{i_{2m}})) \cdot (\varphi_1, \ldots, \varphi_{p-1})$$

for some $l \geq 0$ and $p \in \Omega$. Other cases reduce to this one through the symmetry of the claim (we have an equivalence relation) or, if $\varphi$ is a Boolean combination of formulae, by replacing $\varphi$ by one of its components. Set $\vec{i} = (i_1, \ldots, i_{2m})$ and $\Phi = (\varphi_1, \ldots, \varphi_{p-1})$. Then by assumption on $\varphi$,

$$\mathrm{rk}(\mathrm{fmat}_{\vec{x},\vec{i}}(\Phi, \mathbf{A}, \vec{a})_p) \neq \mathrm{rk}(\mathrm{fmat}_{\vec{x},\vec{i}}(\Phi, \mathbf{B}, \vec{b})_p). \tag{$*$}$$

Spoiler now starts the round by choosing the prime $p$ and picking up pebbles with labels $i_1, \ldots, i_{2m}$. Duplicator has to respond by choosing partitions $\mathbf{P}, \mathbf{Q}$ and a bijection $f : \mathbf{P} \to \mathbf{Q}$, which satisfy the requirements of the game. If Duplicator fails to properly respond to the challenge of Spoiler, then Spoiler wins the game immediately, so assume that $\mathbf{P}, \mathbf{Q}$ and $f$ satisfy the rank condition (**r.c.**). Then the following claim shows that the partitions proposed by Duplicator must contain a block with tuples that disagree on at least one of the $\varphi_i$.

**Claim 1.** *There is a block $P \in \mathbf{P}$ and tuples $\vec{c} \in P$ and $\vec{d} \in f(P)$ for which there is some formula $\varphi_i$ in $\Phi$ such that*

$$\mathbf{A} \models \varphi_i[\vec{a}\tfrac{c_1}{i_1} \cdots \tfrac{c_{2m}}{i_{2m}}] \text{ and } \mathbf{B} \models \neg\varphi_i[\vec{b}\tfrac{d_1}{i_1} \cdots \tfrac{d_{2m}}{i_{2m}}],$$

*or vice versa.*

*Proof of claim.* Suppose, towards a contradiction, that each block $P \in \mathbf{P}$ contains only tuples that all realise one or the other, $\varphi_i$ or $\neg\varphi_i$, and all tuples in $f(P)$ realise the same (corresponding) formula, for each $i \in [p-1]$. Hence, the map $\iota : \mathbf{P} \to \wp([p-1])$ that associates with each $P \in \mathbf{P}$ the set of formulae in $\Phi$ that are realised by some (and hence all) tuples in $P$ is well-defined. Note that for each $P \in \mathbf{P}$, the formulae

$$\bigwedge_{i \in \iota(P)} \varphi_i \text{ and } \bigwedge_{i \in [1, p-1] \setminus \iota(P)} \neg\varphi_i$$

are realised by all tuples in $P$. Now consider the matrix $\mathrm{fmat}_{\vec{x},\vec{i}}(\Phi, \mathbf{A}, \vec{a})_p$ defined over $\mathbf{A}$. By assumption, we can find a labelling $\gamma : \mathbf{P} \to \mathsf{GF}_p$ such that

$$\mathrm{fmat}_{\vec{x},\vec{i}}(\Phi, \mathbf{A}, \vec{a})_p = M_\gamma^{\mathbf{P}} \text{ and } \mathrm{fmat}_{\vec{x},\vec{i}}(\Phi, \mathbf{B}, \vec{b})_p = M_{\gamma \circ f^{-1}}^{\mathbf{Q}}.$$

For instance, $\gamma$ can be defined by $\gamma(P) := \sum_{i \in \iota(P)} i \pmod{p}$ for each $P \in \mathbf{P}$ (as an element of $\mathsf{GF}_p$). But

$$\mathrm{rk}(\mathrm{fmat}_{\vec{x},\vec{i}}(\Phi, \mathbf{A}, \vec{a})_p) \neq \mathrm{rk}(\mathrm{fmat}_{\vec{x},\vec{i}}(\Phi, \mathbf{B}, \vec{b})_p)$$

by (∗), while $\mathrm{rk}(M_\gamma^{\mathbf{P}}) = \mathrm{rk}(M_{\gamma \circ f^{-1}}^{\mathbf{Q}})$ since we assumed that Duplicator's response satisfies condition (**r.c.**). Therefore, we have a contradiction. ◁

Now Spoiler picks some block $P$ that satisfies the statement of the claim. This allows him to place the chosen pebbles on elements $(c_1, \ldots, c_{2m}) \in P$ and $(d_1, \ldots, d_{2m}) \in f(P)$ such that the two structures, with the corresponding pebble placements, can be distinguished by a formula of quantifier rank $q' < q$. □

In the proof of the next lemma, we show that if $(\mathbf{A}, \vec{a}) \equiv^R_{k,m,\Omega} (\mathbf{B}, \vec{b})$, then Duplicator can play one round of the $(k, m, \Omega)$-matrix-equivalence game in a way that ensures that the resulting positions will also be $\equiv^R_{k,m,\Omega}$-equivalent. This gives her a strategy to play the game indefinitely. The idea here is to let Duplicator respond to a challenge of the Spoiler with partitions $\mathbf{P}$ and $\mathbf{Q}$ that are obtained by grouping together in each partition block all the elements realising the same $R^k_{m;\Omega}$-type (with respect to the current game positions). The bijection $f : \mathbf{P} \to \mathbf{Q}$ is similarly defined by pairing together blocks whose elements all realise the same $R^k_{m;\Omega}$-type. We show that if Duplicator plays in this manner, then she can ensure both that condition (**r.c.**) is met and that Spoiler is restricted to placing his pebbles in blocks which do not distinguish the two structures.

**Lemma 3.6.** *If* $(\mathbf{A}, \vec{a}) \equiv^R_{k,m,\Omega} (\mathbf{B}, \vec{b})$ *then Duplicator has a winning strategy in the $(k, m, \Omega)$-matrix-equivalence game starting with positions $(\mathbf{A}, \vec{a})$ and $(\mathbf{B}, \vec{b})$.*

*Proof.* Assume that $(\mathbf{A}, \vec{a}) \equiv^R_{k,m,\Omega} (\mathbf{B}, \vec{b})$ and let $\vec{x} = (x_1, \ldots, x_k)$ be a $k$-tuple of distinct variables. Suppose that Spoiler begins a round of the game by choosing a prime $p \in \Omega$ and picking up pebbles labelled $i_1, \ldots, i_{2m}$, in that sequence. Hereafter, let $\vec{i} = (i_1, \ldots, i_{2m})$. Then we show that Duplicator can respond with partitions that satisfy condition (**r.c.**) and which ensure that all game positions that can result will be $\equiv^R_{k,m,\Omega}$-equivalent. Firstly, for each $\alpha \in \mathrm{Tp}(R^k_{m;\Omega}; \tau, k)$ let

$$P_\alpha := \{\vec{c} \in U(\mathbf{A})^{2m} \mid \mathrm{tp}(R^k_{m;\Omega}; \mathbf{A}, \vec{a}\tfrac{\vec{c}}{\vec{i}}) = \alpha\} \text{ and}$$
$$Q_\alpha := \{\vec{d} \in U(\mathbf{B})^{2m} \mid \mathrm{tp}(R^k_{m;\Omega}; \mathbf{B}, \vec{b}\tfrac{\vec{d}}{\vec{i}}) = \alpha\}.$$

That is, each $P_\alpha$ consists of all $2m$-tuples that, when used to replace elements of $\vec{a}$ according to the index pattern $\vec{i}$, results in a tuple whose type over $\mathbf{A}$ is $\alpha$ (and similarly for each $Q_\alpha$). The strategy of Duplicator is now to respond with

$$\mathbf{P} := \{P_\alpha \mid \alpha \in \Gamma \text{ and } P_\alpha \neq \emptyset\} \quad \text{and} \quad \mathbf{Q} := \{Q_\alpha \mid \alpha \in \Gamma \text{ and } Q_\alpha \neq \emptyset\},$$

where we let $\Gamma := \mathrm{Tp}(R^k_{m;\Omega}; \tau, k)$. To pair the two partitions together, Duplicator gives a mapping $f$ on $\mathbf{P}$ defined by $P_\alpha \mapsto Q_\alpha$ for all non-empty $P_\alpha$. It should be clear that $\mathbf{P}$ and $\mathbf{Q}$ are partitions of $U(\mathbf{A})^m \times U(\mathbf{A})^m$ and $U(\mathbf{B})^m \times U(\mathbf{B})^m$, respectively (just observe that each tuple of elements realises only one type).

We now establish, through a series of claims, that $\mathbf{P}$, $\mathbf{Q}$ and $f$ satisfy the requirements (**r.c.**) of the game; in particular, that $f$ is a bijection $\mathbf{P} \to \mathbf{Q}$.

**Claim 2.** *$f$ is a bijection $\mathbf{P} \to \mathbf{Q}$.*

*Proof of claim.* To prove this claim, it suffices to show that $P_\alpha = \emptyset \Leftrightarrow Q_\alpha = \emptyset$ for all $\alpha \in \text{Tp}(R_{m;\Omega}^k; \tau, k)$. Suppose, towards a contradiction, that there is some $\alpha$ such that $P_\alpha$ is empty while $Q_\alpha$ is not empty (the other case is equivalent, by symmetry of the claim). By the definition of $P_\alpha$, we know that

$$\text{tp}(R_{m;\Omega}^k; \mathbf{A}, \vec{a}\tfrac{\vec{c}}{\vec{i}}) \neq \alpha$$

for all $\vec{c} \in U(\mathbf{A})^{2m}$. This means that for each such tuple $\vec{c}$, there is some formula $\psi_{\vec{c}} \in \alpha$ such that $(\mathbf{A}, \vec{a}\tfrac{\vec{c}}{\vec{i}}) \not\models \psi_{\vec{c}}$. Fix such a formula $\psi_{\vec{c}}$ for each $\vec{c}$ and let $\Psi_\alpha := \bigwedge_{\vec{c}} \psi_{\vec{c}}$; since $\Psi_\alpha$ is defined by conjunction over a finite set of formulae from $\alpha$, it follows that $\Psi_\alpha \in \alpha$. Since $P_\alpha = \emptyset$ it follows that

$$\|\{\vec{c} \in U(\mathbf{A})^{2m} \mid (\mathbf{A}, \vec{a}\tfrac{\vec{c}}{\vec{i}}) \models \Psi_\alpha\}\| = 0. \tag{$\dagger$}$$

This condition can be formalised in $R_{m;\Omega}^k$ as the formula

$$\theta := \text{rk}_p^{=0}((x_{i_1}, \ldots, x_{i_m}), (x_{i_{m+1}}, \ldots, x_{i_{2m}})) . (\Psi_\alpha),$$

for some $p \in \Omega$. This formula asserts that the number of distinct $2m$-tuples $\vec{x}$ that realise $\Psi_\alpha$ over $(\mathbf{A}, \vec{a})$ is nil (more directly, that the matrix defined by $\Psi_\alpha$ is the all-zeroes matrix). By ($\dagger$), it follows that $(\mathbf{A}, \vec{a}) \models \theta$. However, we have $(\mathbf{B}, \vec{b}) \not\models \theta$ since

$$\|\{\vec{d} \in U(\mathbf{B})^{2m} \mid (\mathbf{B}, \vec{b}\tfrac{\vec{d}}{\vec{i}}) \models \Psi_\alpha\}\| > 0,$$

by the assumption $Q_\alpha \neq \emptyset$. Observing that $\theta \in R_{m;\Omega}^k$, we conclude that

$$(\mathbf{A}, \vec{a}) \not\equiv_{k,m,\Omega}^R (\mathbf{B}, \vec{b}),$$

which contradicts the assumption of the lemma. ◁

**Claim 3.** *For each $\alpha$ for which neither $P_\alpha$ nor $Q_\alpha$ are empty, there is a formula $\varphi_\alpha \in R_{m;\Omega}^k$ (depending on both $\mathbf{P}$ and $\mathbf{Q}$) which isolates $P_\alpha$ in $\mathbf{P}$ and $Q_\alpha$ in $\mathbf{Q}$. That is, for all $\vec{c} \in P_\alpha$ and $\vec{d} \in Q_\alpha$,*

$$(\mathbf{A}, \vec{a}\tfrac{\vec{c}}{\vec{i}}) \models \varphi_\alpha \quad and \quad (\mathbf{B}, \vec{b}\tfrac{\vec{d}}{\vec{i}}) \models \varphi_\alpha$$

*while for all $\vec{c} \notin P_\alpha$ and $\vec{d} \notin Q_\alpha$,*

$$(\mathbf{A}, \vec{a}\tfrac{\vec{c}}{\vec{i}}) \not\models \varphi_\alpha \quad and \quad (\mathbf{B}, \vec{b}\tfrac{\vec{d}}{\vec{i}}) \not\models \varphi_\alpha.$$

*Proof of claim.* Let $P_\beta$ be non-empty and let $\vec{c} \in P_\beta$. For each $\alpha \neq \beta$ for which $P_\alpha \neq \emptyset$, fix a formula $\psi_{\vec{c}} \in \alpha$ for which $(\mathbf{A}, \vec{a}\tfrac{\vec{c}}{\vec{i}}) \not\models \psi_{\vec{c}}$. Such a formula must exist, since $\vec{c}$ does not realise the type $\alpha$ over $(\mathbf{A}, \vec{a})$ with respect to the index pattern $\vec{i}$. Doing this for all tuples in $P_\beta$, we let

$$\Psi_{\alpha,\beta}^{\mathbf{P}} := \bigwedge_{\vec{c} \in P_\beta} \psi_{\vec{c}}.$$

By this construction, it follows that $\Psi_{\alpha,\beta}^{\mathbf{P}} \in \alpha$. Now fix a type $\alpha$ with $P_\alpha \neq \emptyset$ and let

$$\varphi_\alpha^{\mathbf{P}} := \bigwedge_{\alpha \neq \beta \text{ and } P_\beta \neq \emptyset} \Psi_{\alpha,\beta}^{\mathbf{P}}.$$

Observe that for all $\vec{c} \in P_\alpha$ with $P_\alpha \neq \emptyset$, we have $(\mathbf{A}, \vec{a}\frac{\vec{c}}{\vec{i}}) \models \varphi_\alpha^{\mathbf{P}}$ and $(\mathbf{A}, \vec{a}\frac{\vec{c}}{\vec{i}}) \not\models \varphi_\beta^{\mathbf{P}}$ for all $\beta \neq \alpha$. In other words, $\varphi_\alpha^{\mathbf{P}}$ isolates $P_\alpha$ in $\mathbf{P}$.

By the same construction, we obtain for each $\alpha$ a formula $\varphi_\alpha^{\mathbf{Q}}$ that isolates $Q_\alpha$ in $\mathbf{Q}$. Then it follows that the formula $\varphi_\alpha := \varphi_\alpha^{\mathbf{P}} \wedge \varphi_\alpha^{\mathbf{Q}} \in \alpha$ isolates both $P_\alpha$ in $\mathbf{P}$ and $Q_\alpha$ in $\mathbf{Q}$. ◁

**Claim 4.** $\operatorname{rk}(M_\gamma^{\mathbf{P}}) = \operatorname{rk}(M_{\gamma \circ f^{-1}}^{\mathbf{Q}})$ for all $\gamma : \mathbf{P} \to \mathsf{GF}_p$.

*Proof of claim.* Without loss of generality, we associate $\mathsf{GF}_p$ with the set $[0, p-1]$ in the canonical way throughout this proof. With that assumption, let $\gamma : \mathbf{P} \to [0, p-1]$ be a labelling. From the definition of $\mathbf{P}$, it can be seen that the collection of blocks labelled $c \in [0, p-1]$ by $\gamma$ corresponds to a finite set of types $\Omega_c \subseteq \operatorname{tp}(R_{m;\Omega}^k; \mathbf{A}, \vec{a})$, with each type $\alpha \in \Omega_c$ isolated by a formula $\varphi_\alpha \in R_{m;\Omega}^k$, by the previous claim. That is, for each type $\alpha$ it holds that

$$\alpha \in \Omega_c \Leftrightarrow \gamma(P_\alpha) = \gamma(\operatorname{ext}_{\vec{i}}^{\varphi_\alpha}(\mathbf{A}, \vec{a})) = c.$$

For $c \in [0, p-1]$, let $\psi_c := \bigvee_{\alpha \in \Omega_c} \varphi_\alpha$; clearly we have $\psi_c \in R_{m;\Omega}^k$. It can now be seen that

$$M_\gamma^{\mathbf{P}} := \sum_{c=0}^{p-1} c \cdot \Big( \sum_{\alpha \in \Omega_c} \operatorname{fmat}_{\vec{x},\vec{i}}(\varphi_\alpha, \mathbf{A}, \vec{a})_p \Big)$$
$$= \sum_{c=1}^{p-1} c \cdot \operatorname{fmat}_{\vec{x},\vec{i}}(\psi_c, \mathbf{A}, \vec{a})_p$$
$$= \operatorname{fmat}_{\vec{x},\vec{i}}(\psi_1, \ldots, \psi_{p-1}, \mathbf{A}, \vec{a})_p,$$

and

$$M_{\gamma \circ f^{-1}}^{\mathbf{Q}} := \sum_{c=1}^{p-1} c \cdot \Big( \sum_{\alpha \in \Omega_c} \operatorname{fmat}_{\vec{x},\vec{i}}(\varphi_\alpha, \mathbf{B}, \vec{b})_p \Big)$$
$$= \sum_{c=1}^{p-1} c \cdot \operatorname{fmat}_{\vec{x},\vec{i}}(\psi_c, \mathbf{B}, \vec{b})_p$$
$$= \operatorname{fmat}_{\vec{x},\vec{i}}(\psi_1, \ldots, \psi_{p-1}, \mathbf{B}, \vec{b})_p.$$

By Lemma 3.4 we know that

$$\operatorname{rk}\big(\operatorname{fmat}_{\vec{x},\vec{i}}(\psi_1, \ldots, \psi_{p-1}, \mathbf{A}, \vec{a})_p\big) = \operatorname{rk}\big(\operatorname{fmat}_{\vec{x},\vec{i}}(\psi_1, \ldots, \psi_{p-1}, \mathbf{B}, \vec{b})_p\big).$$

Hence, $\operatorname{rk}(M_\gamma^{\mathbf{P}}) = \operatorname{rk}(M_{\gamma \circ f^{-1}}^{\mathbf{Q}})$ over $\mathsf{GF}_p$, as required. ◁

By the three preceding claims, it can be seen that for any block $P \in \mathbf{P}$, any choice of elements $(c_1, \ldots, c_{2m}) \in P$ and $(d_1, \ldots, d_{2m}) \in f(P)$ that Spoiler can make will result in tuples $\vec{a}\frac{\vec{c}}{\vec{i}}$ and $\vec{b}\frac{\vec{d}}{\vec{i}}$ that realise the same $R_{m;\Omega}^k$-type. This shows that Spoiler has a strategy to maintain $\equiv_{k,m,\Omega}^R$-equivalence of game positions, which completes the proof of the lemma. □

# 4 Playing with invertible linear maps

It follows from the pebble-game characterisation of finite-variable counting logics that the equivalence $\equiv_k^C$ on finite structures is decidable in polynomial time. Essentially, this is because the number of possible moves for Duplicator at any particular stage of the game can be inductively combined into a structural invariant that completely characterises the game equivalence, and this invariant can be constructed in polynomial time (see Otto [16] for details). In light of Theorem 3.1, we may therefore ask whether the matrix-equivalence game for $R_{m;\Omega}^k$ can be used to give a similar result for equivalence in finite-variable rank logic; that is, whether we can decide $\equiv_{k,m,\Omega}^R$ on finite structures in polynomial time. Unfortunately, unlike the case with the counting game, there does not seem to be an effective way to encode complete information about a winning strategy in the matrix-equivalence game into a polynomial-size invariant. The main problem is the game condition (**r.c.**); this requires Duplicator to show that each pair of matrices $M_\gamma^{\mathbf{P}}$ and $M_{\gamma \circ f^{-1}}^{\mathbf{Q}}$ is equivalent (that is, have the same rank) but the number of these matrices is *exponential* in the size of the partition.

In an attempt to avoid this exponential number of matrix combinations, we define a modification of the matrix-equivalence game which is based on invertible linear maps. In this game, Duplicator is required to specify a bijection between the partitions of the two game structures as the conjugacy action of a single invertible matrix. In that sense, the invertible-map game can be seen as the natural extension of the bijection game for counting logics, where we replace bijections with invertible maps. In [5] it had been asked whether such a game might characterise definability in finite-variable rank logic. We show that equivalence in the invertible-map game does in fact *refine* the relations $\equiv_{k,m,\Omega}^R$ while it is not known whether the converse holds. We also establish that equivalence in the invertible-map game can be decided in polynomial time, which is not known to be true for the $\equiv_{k,m,\Omega}^R$ as we discussed above. We see one application of this new game equivalence in the next section, where we define algorithms for testing graph isomorphism by playing the invertible-map game on finite graphs.

## 4.1 Invertible-map game

Let $k$ and $m$ be positive integers with $2m \leq k$ and let $\Omega$ be a finite and non-empty set of primes. The game board of the $k$-pebble $m$-ary *invertible-map game* over $\Omega$ (or $(k, m, \Omega)$-invertible-map game for short) consists of two structures $\mathbf{A}$ and $\mathbf{B}$ of the same vocabulary, each with $k$ pebbles labelled $1, \ldots, k$ (and initial placement of pebbles $\vec{a}$ over $\mathbf{A}$ and $\vec{b}$ over $\mathbf{B}$, as before). If $\|\mathbf{A}\| \neq \|\mathbf{B}\|$ or the mapping defined by the initial pebble positions is not a partial isomorphism, then Spoiler wins the game immediately. Otherwise, each round of the game is played as follows.

1. Spoiler chooses a prime $p \in \Omega$ and picks up $2m$ pebbles in some order from $\mathbf{A}$ and the $2m$ corresponding pebbles in the same order from $\mathbf{B}$.

2. Duplicator has to respond by choosing
   - a partition $\mathbf{P}$ of $U(\mathbf{A})^m \times U(\mathbf{A})^m$,
   - a partition $\mathbf{Q}$ of $U(\mathbf{B})^m \times U(\mathbf{B})^m$, with $\|\mathbf{P}\| = \|\mathbf{Q}\|$, and

- a non-singular $U(\mathbf{B})^m \times U(\mathbf{A})^m$ matrix $S$ over $\mathsf{GF}_p$,

for which it holds that the map $f : \mathbf{P} \to \mathbf{Q}$ defined by

$$P \mapsto Q \quad \text{iff} \quad S \cdot \chi_P \cdot S^{-1} = \chi_Q \qquad (\circledast)$$

is *total* and *bijective*, where we view $\chi_P$ and $\chi_Q$ as $\{0,1\}$-matrices over $\mathsf{GF}_p$.

3. Spoiler next chooses a block $P \in \mathbf{P}$ and places the $2m$ chosen pebbles from $\mathbf{A}$ on the elements of some tuple in $P$ (in the order they were chosen earlier) and the corresponding $2m$ pebbles from $\mathbf{B}$ on the elements of some tuple in $f(P)$.

This completes one round in the game. If, after this exchange, the partial map from $\mathbf{A}$ to $\mathbf{B}$ defined by the pebbled positions is not a partial isomorphism, or if Duplicator is unable to produce the necessary triple $(\mathbf{P}, \mathbf{Q}, S)$, then Spoiler has won the game; otherwise it can continue for another round.

We write $(\mathbf{A}, \vec{b}) \approx_{m,\Omega}^{k} (\mathbf{B}, \vec{b})$ to denote that Duplicator has a strategy to play forever in the $(k, m, \Omega)$-invertible-map game with starting positions $(\mathbf{A}, \vec{a})$ and $(\mathbf{B}, \vec{b})$. Clearly, in the matrix-equivalence game it is sufficient for Duplicator to demonstrate the existence of a single similarity transformation that relates all linear combinations of partition matrices, since similar matrices have the same rank. Hence, we establish that $\approx_{m,\Omega}^{k}$ refines $\equiv_{k,m,\Omega}^{R}$ for all values of $k$, $m$ and $\Omega$.

**Lemma 4.1.** *Duplicator has a winning strategy in the $(k, m, \Omega)$-matrix-equivalence game starting on $(\mathbf{A}, \vec{a})$ and $(\mathbf{B}, \vec{b})$ if she has a winning strategy in the $(k, m, \Omega)$-invertible-map game starting on $(\mathbf{A}, \vec{a})$ and $(\mathbf{B}, \vec{b})$.*

### 4.2 Analysis of winning strategies

In this section we look more closely at the type of response that can be given by Duplicator in a winning strategy in the invertible-map game. More specifically, we consider some basic structural properties of the partitions $\mathbf{P}$ and $\mathbf{Q}$ and the invertible maps $S$ that Duplicator produces during the game play. In order to state our results, we first need to establish some notation.

Let $m \in \mathbb{N}$ and consider $m$-tuples $\vec{x}, \vec{y} \in A^m$ for some non-empty set $A$. We say that $\vec{x}$ and $\vec{y}$ are *equality equivalent*, written $\vec{x} \equiv_m \vec{y}$, if for all $i, j \in [m]$ it holds that $x_i = x_j$ if and only if $y_i = y_j$. We call the equivalence classes of $\equiv_m$ *equality types*. Let $T_m(A)$ be the set of all equality types on $m$-tuples over $A$; that is, $T_m A := A^m / \equiv_m$ is the partition of $A^m$ into equality types. For $\vec{x} \in A^m$, we write $\mathbf{eqtp}_m(\vec{x})$ to denote the $\equiv_m$-equivalence class of $\vec{x}$. We say that $\vec{x}$ has equality type $\alpha \in T_m(A)$ if $\mathbf{eqtp}_m(\vec{x}) = \alpha$.

**Lemma 4.2.** *Let $(\mathbf{A}, \vec{a}) \approx_{m,\Omega}^{k} (\mathbf{B}, \vec{b})$ and suppose that at as a part of her winning strategy in the invertible-map game starting on $(\mathbf{A}, \vec{a})$ and $(\mathbf{B}, \vec{b})$, Duplicator responds to a challenge of Spoiler by giving partitions $\mathbf{P}$ and $\mathbf{Q}$ of $U(\mathbf{A})^m \times U(\mathbf{A})^m$ and $U(\mathbf{B})^m \times U(\mathbf{B})^m$, respectively, and an invertible matrix $S$. Then for each $P \in \mathbf{P}$ and all $\vec{a} \in P$ and $\vec{b} \in f(P)$, it holds that $\mathbf{eqtp}_{2m}(\vec{a}) = \mathbf{eqtp}_{2m}(\vec{b})$, where $f : \mathbf{P} \to \mathbf{Q}$ is the bijection induced by $S$.*

*Proof.* Suppose, towards a contradiction, that there is a block $P \in \mathbf{P}$ and tuples $\vec{a} \in P$ and $\vec{b} \in f(P)$ with $\mathbf{eqtp}_{2m}(\vec{a}) \neq \mathbf{eqtp}_{2m}(\vec{b})$. Since Spoiler is free to choose where to put down his pebbles within matching blocks after Duplicator has given the partitions $\mathbf{P}$ and $\mathbf{Q}$, he can exploit this fact to immediately win the game. More specifically, Spoiler's strategy is to choose the block $P$ and to place the chosen pebbles over $\mathbf{A}$ on the elements of $\vec{a}$ and the corresponding pebbles over $\mathbf{B}$ on the elements of $\vec{b}$. This guarantees that at the end of the round, the mapping between the two structures given by the pebbled elements is not a partial isomorphism which means that Spoiler wins the game, contradicting the assumption that $\mathbf{P}$ and $\mathbf{Q}$ were played as a part of a winning strategy for Duplicator. □

An immediate corollary of this lemma is that any partition played by Duplicator as a part of a winning strategy must refine the basic equality-type partition of $2m$-tuples.

**Corollary 4.3** (Partitions in a winning strategy). *If $(\mathbf{A}, \vec{a}) \approx_{m,\Omega}^{k} (\mathbf{B}, \vec{b})$ then $\mathbf{P} \subseteq T_{2m}(U(\mathbf{A}))$ and $\mathbf{Q} \subseteq T_{2m}(U(\mathbf{B}))$ for any partitions $\mathbf{P}$ and $\mathbf{Q}$ that Duplicator plays as a part of her winning strategy.*

The next result states that in order to win the game, Duplicator can always play by offering invertible matrices that can be block-decomposed into a direct sum of smaller matrices, according to the partition of the rows and columns into equality types. To explain this further, fix an enumeration $\gamma_1, \ldots, \gamma_N$ of the equality types in $T_m(U(\mathbf{A}) \cup U(\mathbf{B}))$ and write $\alpha_i := \gamma_i \cap U(\mathbf{A})^m$ and $\beta_i := \gamma_i \cap U(\mathbf{B})^m$ for the restriction of $\gamma_i$ to tuples over $U(\mathbf{A})$ and $U(\mathbf{B})$, respectively. Then by this next lemma, we know that if Duplicator has a winning strategy in the invertible-map game, then she can always play by giving invertible matrices $S$ of the following kind

$$ S = \bigoplus_{i=1}^{N} S_i = \begin{pmatrix} \alpha_1 & \alpha_2 & \cdots & \alpha_N \\ S_1 & & & \\ & S_2 & & \\ & & \ddots & \\ & & & S_N \end{pmatrix} \begin{matrix} \beta_1 \\ \beta_2 \\ \vdots \\ \beta_N \end{matrix}, $$

where each $S_i$ is an invertible $\beta_i \times \alpha_i$ matrix. This implies, in particular, that for each $i \in \{1, \ldots, N\}$, we have $S^{-1} = \bigoplus_{i=1}^{N} S_i^{-1}$.

**Lemma 4.4.** *Suppose $(\mathbf{A}, \vec{b}) \approx_{m,\Omega}^{k} (\mathbf{B}, \vec{b})$ and write $\Gamma := T_m(U(\mathbf{A}) \cup U(\mathbf{B}))$ to denote the set of all equality types of $m$-tuples over $U(\mathbf{A})$ and $U(\mathbf{B})$. Then Duplicator has a winning strategy that allows her to respond to any move by the Spoiler in the invertible-map game starting on $(\mathbf{A}, \vec{b})$ and $(\mathbf{B}, \vec{b})$ with a triple $(\mathbf{P}, \mathbf{Q}, S)$ where $S = \bigoplus_{\alpha \in \Gamma} S_\alpha$ is a block diagonal matrix with each $S_\alpha$ a non-singular matrix indexed by $\alpha \cap U(\mathbf{B})^m \times \alpha \cap U(\mathbf{A})^m$.*

*Proof.* As above, fix an enumeration $\gamma_1, \ldots, \gamma_N$ of the equality types in $T_m(U(\mathbf{A}) \cup U(\mathbf{B}))$ and write $\alpha_i := \gamma_i \cap U(\mathbf{A})^m$ and $\beta_i := \gamma_i \cap U(\mathbf{B})^m$ for the restriction of $\gamma_i$ to tuples over $U(\mathbf{A})$ and $U(\mathbf{B})$, respectively. Assuming $(\mathbf{A}, \vec{b}) \approx_{m,\Omega}^{k} (\mathbf{B}, \vec{b})$, suppose that, as as part of her winning strategy, Duplicator responds to a challenge of Spoiler by giving a triple $(\mathbf{P}, \mathbf{Q}, S)$. Partitioning the row and column sets of $S$ according to equality types, we can write the matrices $S$ and $S^{-1}$ in block form as

$$S = \begin{pmatrix} & \alpha_1 & \alpha_2 & \ldots & \alpha_N \\ \beta_1 & S_{11} & S_{12} & \ldots & S_{1N} \\ \beta_2 & S_{21} & S_{22} & \ldots & S_{2N} \\ \vdots & \vdots & \vdots & \ddots & \vdots \\ \beta_N & S_{N1} & S_{N2} & \ldots & S_{NN} \end{pmatrix} \text{ and}$$

$$S^{-1} = \begin{pmatrix} & \beta_1 & \beta_2 & \ldots & \beta_N \\ \alpha_1 & T_{11} & T_{12} & \ldots & T_{1N} \\ \alpha_2 & T_{21} & T_{22} & \ldots & T_{2N} \\ \vdots & \vdots & \vdots & \ddots & \vdots \\ \alpha_N & T_{N1} & T_{N2} & \ldots & T_{NN} \end{pmatrix}.$$

Observe that while $S$ is invertible, that of course does not imply that any of the block matrices $S_{ij}$ are themselves invertible.

Now consider some equality type $\alpha_i$. It can be seen that there is some set $\mathbf{P'} \subseteq \mathbf{P}$ of blocks in $\mathbf{P}$ for which it holds that $M_i := \sum_{P \in \mathbf{P'}} \chi_P$ is the direct sum of the identity matrix on $\alpha_i$ and all-zeroes matrices. That is,

$$M = \begin{pmatrix} & \alpha_1 & \ldots & \alpha_{i-1} & \alpha_i & \alpha_{i+1} & \ldots & \alpha_N \\ \alpha_1 & 0 & \ldots & 0 & 0 & 0 & \ldots & 0 \\ \vdots & \vdots & \ddots & \vdots & \vdots & \vdots & \ddots & \vdots \\ \alpha_{i-1} & 0 & \ldots & 0 & 0 & 0 & \ldots & 0 \\ \alpha_i & 0 & \ldots & 0 & I & 0 & \ldots & 0 \\ \alpha_{i+1} & 0 & \ldots & 0 & 0 & 0 & \ldots & 0 \\ \vdots & \vdots & \ddots & \vdots & \vdots & \vdots & \ddots & \vdots \\ \alpha_N & 0 & \ldots & 0 & 0 & 0 & \ldots & 0 \end{pmatrix}.$$

Similarly, it can be seen that $S \cdot M \cdot S^{-1} = \sum_{P \in \mathbf{P'}} f(\chi_P)$ is the direct sum of the identity matrix on $\beta_i$ and all-zeroes matrices. By Lemma 4.2, it follows that all entries of $S \cdot M \cdot S^{-1}$ outside the $\beta_i \times \beta_i$ sub-matrix have to be zero. Hence, expanding the product $S \cdot M \cdot S^{-1}$, we see that

$$S \cdot M \cdot S^{-1} = \begin{pmatrix} & \beta_1 & \ldots & \beta_{i-1} & \beta_i & \beta_{i+1} & \ldots & \beta_N \\ \beta_1 & 0 & \ldots & 0 & 0 & 0 & \ldots & 0 \\ \vdots & \vdots & \ddots & \vdots & \vdots & \vdots & \ddots & \vdots \\ \beta_{i-1} & 0 & \ldots & 0 & 0 & 0 & \ldots & 0 \\ \beta_i & 0 & \ldots & 0 & S_{ii} \cdot I \cdot T_{ii} & 0 & \ldots & 0 \\ \beta_{i+1} & 0 & \ldots & 0 & 0 & 0 & \ldots & 0 \\ \vdots & \vdots & \ddots & \vdots & \vdots & \vdots & \ddots & \vdots \\ \beta_N & 0 & \ldots & 0 & 0 & 0 & \ldots & 0 \end{pmatrix},$$

which shows that $S_{ii} \cdot T_{ii} = I$; that is, $S_{ii}$ is invertible with inverse $S_{ii}^{-1} = T_{ii}$. Now by the result of Lemma 4.2, we know that any $P \in \mathbf{P}$ will be within an $\alpha_i \times \alpha_j$ sub-matrix, for some $i$ and $j$, and that correspondingly $f(P)$ will be within a $\beta_i \times \beta_j$ sub-matrix. Considering the block form of $S$ we expressed above, it follows that the part of $f(P)$ falling within the $\beta_i \times \beta_j$ sub-matrix is $S_{ii} \cdot \chi_P \cdot S_{jj}^{-1}$. Therefore, writing $R_i := R_{ii}$, it can be seen that by taking $R := \bigoplus_{i=1}^{N} R_i$, we get a block-diagonal invertible matrix which, along with the partitions **P** and **Q**, satisfies condition (✻) of the game and which preserves the winning strategy of the Duplicator. □

Finally, the following lemma shows that with increasing $k$, $m$ or $\Omega$, we get a decreasing chain of equivalence relations on $\text{fin}[\tau; k]$.

**Lemma 4.5.** *For all $k, m, p \in \mathbb{N}$, with $2m \leq k$ and $p$ prime, and all finite sets of primes $\Omega$, it holds that $\approx_{m,\Omega}^{k+1} \subseteq \approx_{m,\Omega}^{k}$, $\approx_{m,\Omega \cup \{p\}}^{k} \subseteq \approx_{m,\Omega}^{k}$ and $\approx_{m+1,\Omega}^{k} \subseteq \approx_{m,\Omega}^{k}$.*

*Proof.* The first two inclusions of the lemma follow trivially from the definition of the game. To establish the last inclusion, we suppose that Duplicator has a winning strategy in the $(k, m+1, \Omega)$-invertible-map game on $(\mathbf{A}, \vec{a})$ and $(\mathbf{B}, \vec{b})$ and use that to construct for her a winning strategy in the $(k, m, \Omega)$-game on the same game board. The new strategy is obtained by analysing at each round of the $(k, m, \Omega)$-game the response that would be used by Duplicator in her winning strategy in the $(k, m+1, \Omega)$-game. More specifically, we simulate one move of the latter game by making the "$(k, m+1, \Omega)$-Spoiler" make the same move as he would make in the former game. Suppose that by this simulation, Duplicator will respond to Spoiler's challenge by playing $(\mathbf{P}, \mathbf{Q}, S)$ satisfying condition (✻) of the invertible-map game, where **P** and **Q** are partitions of $U(\mathbf{A})^{m+1} \times U(\mathbf{A})^{m+1}$ and $U(\mathbf{B})^{m+1} \times U(\mathbf{B})^{m+1}$, respectively. Let $X_m \subseteq U(\mathbf{A})^{m+1}$ and $Y_m \subseteq U(\mathbf{B})^{m+1}$ denote the sets of all $(m+1)$-tuples of elements from $U(\mathbf{A})$ and $U(\mathbf{B})$, respectively, whose first two components are equal. Then it follows from Corollary 4.3 that there are $\mathbf{P}' \subseteq \mathbf{P}$ and $\mathbf{Q}' \subseteq \mathbf{Q}$ that give partitions of $X_m$ and $Y_m$, respectively. We get the desired partitions $\mathbf{P}''$ and $\mathbf{Q}''$ for Duplicator in the $(k, m, \Omega)$-game by taking $\mathbf{P}''$ and $\mathbf{Q}''$ to be the projections of $\mathbf{P}'$ and $\mathbf{Q}'$ onto the last $m$ components.

By Lemma 4.4, we can assume without loss of generality that the matrix $S$ has the form $S = \bigoplus_{\alpha \in \Gamma} S_\alpha$, where $\Gamma$ is the set of all equality types on $(m+1)$-tuples realised over $U(\mathbf{A})$ and $U(\mathbf{B})$. Let $\Gamma' \subseteq \Gamma$ denote the set of all equality types in $\Gamma$ satisfying the condition "the first and second components are equal", and let $S' := \bigoplus_{\alpha \in \Gamma'} S_\alpha$. Let $S''$ denote the invertible matrix obtained from $S'$ after projecting the sets indexing the rows and columns of $S'$ onto their last $m$ components; that is, $S''$ is an $U(\mathbf{B})^m \times U(\mathbf{A})^m$ matrix while $S'$ is an $U(\mathbf{B})^{m+1} \times U(\mathbf{A})^{m+1}$ matrix.

Now it is straightforward to verify, based on the assumption that $\mathbf{P}, \mathbf{Q}$ and $S$ were played as a part of a winning strategy, that the triple $(\mathbf{P}'', \mathbf{Q}'', S'')$ satisfies condition (✻) of the invertible-map game and that any move that Spoiler makes subsequently will result in a game position from where Duplicator can also play forever in the $(k, m+1, \Omega)$-game. This inductively gives her a strategy to play forever in the $(k, m, \Omega)$-game, as claimed. □

### 4.3 Complexity of the game equivalence

In this section we show that for each $k$ and vocabulary $\tau$, there is an algorithm that decides whether $(\mathbf{A}, \vec{a}) \approx_{m,\Omega}^{k} (\mathbf{B}, \vec{b})$ in time polynomial in $np_{\max}$, where $n$ is the size of both **A** and **B** and $p_{\max}$ is the largest prime in $\Omega$. To simplify our notation, fix $k$, $m$, $\Omega$ and $\tau$. In order to analyse the structure of the game equivalence, we consider a stratification of $\approx_{m,\Omega}^k$ by the number of rounds in the game. More specifically, we let $\sim_i$ be the binary relation on $\text{fin}[\tau; k]$ defined by $(\mathbf{A}, \vec{a}) \sim_i (\mathbf{B}, \vec{b})$ if Duplicator has a strategy to play for up to $i$ rounds in the $(k, m, \Omega)$-invertible-map game on $(\mathbf{A}, \vec{a})$ and $(\mathbf{B}, \vec{b})$. This relation can be characterised inductively as follows, where we write $\text{atp}(\mathbf{A}, \vec{a})$ to denote the atomic type of $\vec{a}$ in $\mathbf{A}$.

**Lemma 4.6.** *For all $(\mathbf{A}, \vec{a}), (\mathbf{B}, \vec{b}) \in \text{fin}[\tau; k]$ we have*

$(\mathbf{A}, \vec{a}) \sim_0 (\mathbf{B}, \vec{b})$ *iff* $\text{atp}(\mathbf{A}, \vec{a}) = \text{atp}(\mathbf{B}, \vec{b})$

$(\mathbf{A}, \vec{a}) \sim_{i+1} (\mathbf{B}, \vec{b})$ *iff* $(\mathbf{A}, \vec{a}) \sim_i (\mathbf{B}, \vec{b})$ *and for all $p \in \Omega$ and all $\vec{j} \in [k]^{2m}$ with distinct values there is an invertible matrix $S$ over $\mathsf{GF}_p$ such that for all $\alpha \in \text{fin}[\tau; k]/\sim_i$:*
$\underbrace{S \cdot \text{extmat}_{\vec{j}}^\alpha(\mathbf{A}, \vec{a}) \cdot S^{-1} = \text{extmat}_{\vec{j}}^\alpha(\mathbf{B}, \vec{b}).}_{(\star\star)}$

For the proof of the inductive step of Lemma 4.6, the "if" direction is fairly straightforward (it essentially specifies a sufficient response for Duplicator in one round of the game). For the converse, it needs to be shown that any partition played by Duplicator as a part of an $(i + 1)$-round winning strategy has to be a refinement of the partition of the corresponding structure into $\sim_i$-equivalence classes. Combining Lemma 4.6 with the fact that the matrix-similarity relation is transitive, we get the following result.

**Corollary 4.7.** $\sim_i$ *is an equivalence relation on $\text{fin}[\tau; k]$ for each $i \in \mathbb{N}_0$.*

*Proof.* We prove this by induction on $i$, noting that the base case $\sim_0$ follows directly from the characterisation of Lemma 4.6. Suppose that it has been shown that $\sim_i$ is an equivalence relation for some $i$. By Lemma 4.6, it then follows that $\sim_{i+1}$ is both reflexive and symmetric. To show transitivity, suppose that $(\mathbf{A}, \vec{a}) \sim_{i+1} (\mathbf{B}, \vec{b})$ and $(\mathbf{B}, \vec{b}) \sim_{i+1} (\mathbf{C}, \vec{c})$. Consider some $\vec{i}, \vec{j} \in [k]^m$ and let $S$ and $T$ be nonsingular $U(\mathbf{B})^m \times U(\mathbf{A})^m$ and $U(\mathbf{C})^m \times U(\mathbf{B})^m$ matrices, respectively, witnessing the similarity condition of the previous lemma. That is, for all $\alpha \in \text{fin}[\tau; k]/\sim_i$, it holds that

$$S \cdot \text{extmat}_{\vec{i}}^\alpha(\mathbf{A}, \vec{a}) \cdot S^{-1} = \text{extmat}_{\vec{i}}^\alpha(\mathbf{B}, \vec{b}) \text{ and}$$
$$T \cdot \text{extmat}_{\vec{i}}^\alpha(\mathbf{B}, \vec{b}) \cdot T^{-1} = \text{extmat}_{\vec{i}}^\alpha(\mathbf{C}, \vec{c}).$$

But then the $U(\mathbf{C})^m \times U(\mathbf{A})^m$ matrix $T \cdot S$ is nonsingular and satisfies

$$(T \cdot S) \cdot \text{extmat}_{\vec{i}}^\alpha(\mathbf{A}, \vec{a}) \cdot (T \cdot S)^{-1} = \text{extmat}_{\vec{i}}^\alpha(\mathbf{C}, \vec{c})$$

for all $\alpha \in \text{fin}[\tau; k]/\sim_i$. □

In particular, since $\approx_{m,\Omega}^k$ coincides with the intersection of the $\sim_i$ over all $i$, it follows that $\approx_{m,\Omega}^k$ is an equivalence relation on $\text{fin}[\tau; k]$.

Now consider some $(\mathbf{A}, \vec{a})$ and $(\mathbf{B}, \vec{b})$ in $\text{fin}[\tau; k]$ and assume that $\|\mathbf{A}\| = \|\mathbf{B}\| = n$. Since the number of distinct positions in the game starting with $(\mathbf{A}, \vec{a})$ and $(\mathbf{B}, \vec{b})$ is bounded by a polynomial in $n$, it follows that there is some polynomial

$q : \mathbb{N} \to \mathbb{N}$ (depending only on $k$ and $\tau$) such that if Duplicator can play the game for at least $q(n)$ rounds, then she has a strategy to play forever. In other words, $(\mathbf{A}, \vec{a}) \approx_{m,\Omega}^{k} (\mathbf{B}, \vec{b})$ if and only if $(\mathbf{A}, \vec{a}) \sim_{q(n)} (\mathbf{B}, \vec{b})$. To decide $(\mathbf{A}, \vec{a}) \approx_{m,\Omega}^{k} (\mathbf{B}, \vec{b})$, we inductively construct the graph of $\approx_{m,\Omega}^{k}$, restricted to $\mathbf{A}$ and $\mathbf{B}$, as follows. Initially, we partition the elements of $U(\mathbf{A})^k \dot\cup U(\mathbf{B})^k$ by their atomic equivalance, which is just $\sim_0$. For the induction step, suppose we have constructed $\sim_i$. Then to compute the refinement $\sim_{i+1}$, we consider each $\sim_i$-equivalent pair $(\vec{c}, \vec{d})$ and check whether condition $(\star\star)$ of Lemma 4.6 is satisfied. That is, for each $p \in \Omega$ and $\vec{j} \in [k]^{2m}$, we let $\mathcal{C} = (C_\alpha)$ and $\mathcal{D} = (D_\alpha)$ be the families of extension matrices defined by $\vec{j}$ over $\vec{c}$ and $\vec{d}$, respectively, indexed by all equivalence classes $\alpha$ of $\sim_i$ (where $C_\alpha = \mathrm{extmat}_{\vec{j}}^{\alpha}(\mathbf{A}, \vec{c})$ if $\vec{c}$ is defined over $\mathbf{A}$, and similarly for $D_\alpha$). Here it is important to note that it suffices to consider only equivalence classes of $\sim_i$ *restricted to* $\mathbf{A}$ *and* $\mathbf{B}$. Therefore, the number of extension matrices that we need to consider is bounded by a polynomial in $n$.

At this stage it remains to determine whether the pair of matrix tuples $\mathcal{C}$ and $\mathcal{D}$ are *simultaneously similar*: that is, whether there is a non-singular matrix $S$ such that $S \cdot C_\alpha \cdot S^{-1} = D_\alpha$ for all $C_\alpha \in \mathcal{C}$. By a result of Chistov et al. [3], this problem is in polynomial time over all finite fields.

**Proposition 4.1** (Chistov, Karpinsky and Ivanyov)**.** *There is a deterministic algorithm that, given two families of $N \times N$ matrices $\mathcal{C} = (C_1, \ldots, C_l)$ and $\mathcal{D} = (D_1, \ldots, D_l)$ over a finite field $\mathsf{GF}_q$, determines in time $\mathrm{poly}(N, l, q)$ whether $\mathcal{C}$ and $\mathcal{D}$ are simultaneously similar.*

By our discussion above, it follows that we can construct the graph of $\approx_{m,\Omega}^{k}$ restricted to $\mathbf{A}$ and $\mathbf{B}$ in a polynomial number of steps. At each step, we need to check a polynomial number of matrix tuples for simultaneous similarity, where each tuple has polynomial length. This gives us a proof of the following theorem.

**Theorem 4.8.** *For each $\tau$ there is a deterministic algorithm that, given $(\mathbf{A}, \vec{a})$, $(\mathbf{B}, \vec{b}) \in \mathrm{fin}[\tau; k]$ (with $\|\mathbf{A}\| = \|\mathbf{B}\| = n$), $m \in \mathbb{N}$ with $2m \leq k$ and a finite set of primes $\Omega$, decides whether $(\mathbf{A}, \vec{a}) \approx_{m,\Omega}^{k} (\mathbf{B}, \vec{b})$ in time $(np)^{\mathcal{O}(k)}$ where $p$ is the largest prime in $\Omega$.*

Observe that this implies that for each fixed $k$, we can decide $\approx_{m,\Omega}^{k}$ in polynomial time, where $\Omega$ can be a part of the input and $m \leq k$.

# 5 Application to the graph isomorphism problem

By considering the invertible-map game equivalence $\approx_{m,\Omega}^{k}$ on the class of all finite graphs, we get a family of polynomial-time algorithms for stratifying the graph isomorphism relation. More specifically, for each $k$, $m$ and $\Omega$, we write $\mathrm{IM}_{m,\Omega}^{k}$ to denote the following algorithm on a pair of finite graphs $\mathbf{G}$ and $\mathbf{H}$:

> If $\|\mathbf{G}\| \neq \|\mathbf{H}\|$ then output "not isomorphic". Otherwise, compute $\approx_{m,\Omega}^{k}$ (restricted to $\mathbf{G}$ and $\mathbf{H}$) on the set $U(\mathbf{G})^k \dot\cup U(\mathbf{H})^k$ by applying the algorithm of Theorem 4.8 for all tuples in $U(\mathbf{G})^k \dot\cup U(\mathbf{H})^k$. If the result is that there is some equivalence class $\alpha$ of $\approx_{m,\Omega}^k$ such that $\|\alpha \cap U(\mathbf{G})^k\| \neq \|\alpha \cap U(\mathbf{H})^k\|$ then output "not isomorphic"; else output "isomorphic".

It follows from Theorem 4.8 that $\mathrm{IM}_{m,\Omega}^k$ runs in polynomial time for a fixed $k$. While the algorithm will always correctly identify isomorphic graphs, it may fail to distinguish between non-isomorphic instances. Furthermore, it can be seen that for each pair of graphs, there is always a value of $k$ for which $\mathrm{IM}_{m,\Omega}^k$ correctly determines isomorphism for any $m$ and any finite set of primes $\Omega$.

The procedure for $\mathrm{IM}_{m,\Omega}^k$ that we outlined above bears a strong resemblance to the well-known Weisfeiler-Lehman method for graph isomorphism (see [2] for a description of the method). It was shown by Cai, Fürer and Immerman [2] that Duplicator has a winning strategy in the $(k+1)$-pebble bijection game on $\mathbf{G}$ and $\mathbf{H}$ if and only if $\mathbf{G}$ and $\mathbf{H}$ are not distinguished by the $k$-dimensional Weisfeiler-Lehman algorithm ($\mathrm{WL}^k$). Combining this characterisation of the Weisfeiler-Lehman algorithm with lemmas 3.3 and 4.1, we have that

$\mathbf{G}$ and $\mathbf{H}$ are distinguished by $\mathrm{WL}^k$
$\Rightarrow \mathbf{G}$ and $\mathbf{H}$ are distinguished by $\mathrm{IM}_{m,\Omega}^{k+2m}$ for *all* $m$ and prime sets $\Omega$.

In [2], Cai et al. showed how to construct for each $k \in \mathbb{N}$ a pair of non-isomorphic graphs (named "CFI graphs") that are indistinguishable in $\mathrm{WL}^k$. Later, it was shown by Dawar et al. [4] that there is a fixed sentence of first-order logic with rank operators over $\mathsf{GF}_2$ that can distinguish between any pair of these CFI graphs. This construction was extended by Holm [12], who showed that for any prime $p$, there are families of non-isomorphic graphs that can be distinguished by first-order logic with rank operators over $\mathsf{GF}_p$ but not by any fixed dimension of Weisfeiler-Lehman. Hence, it follows that the family of $\mathrm{IM}_{m,\Omega}^k$ algorithms provide a way of stratifying the graph isomorphism relation which goes beyond that given by the Weisfeiler-Lehman algorithms.

**Proposition 5.1.** *For each prime $p$ and $k \geq 1$, there is a pair of non-isomorphic graphs $\mathbf{G}$ and $\mathbf{H}$ that can be distinguished by $\mathrm{IM}_{\{p\},1}^3$ but not by $\mathrm{WL}^k$.*

Finally, we remark that Derksen [6] has recently described a family of polynomial-time algorithms that also give an approximation to graph isomorphism that goes beyond that of the Weisfeiler-Lehman method. While Derksen's method partly builds on the simultaneous-similarity algorithm of Chistov et al. [3] (Proposition 4.1), it also draws heavily on techniques from algebraic geometry and category theory and seems very different from the game-based approach that we describe. Nevertheless, it is an open problem whether these two approaches can be related.

## 6 Discussion

A natural question that is raised by the definitions of the games we have presented in this paper, is how to use them to establish inexpressibility results. A step in this direction is presented in [12] where it is shown that for any prime $p$, there is a property definable in first-order logic with rank operators over $\mathsf{GF}_p$ which is not closed under $\equiv_{k,1,\{q\}}^R$ for any $k$ and primes $q \neq p$. This method can be further extended to work for all sets of primes $\Omega \neq \Gamma$ rather than just single primes. It would be interesting to lift this up to arities higher than 1, but playing the game poses combinatorial difficulties.

Another interesting direction would be to establish the precise relationship between the two games we consider. While we showed that the invertible-map game gives a refinement of the matrix-equivalence game (that is, a winning strategy for Duplicator in the former gives a winning strategy in the latter), it is not known whether this refinement is strict. Might it be the case that for any $k$ and $m$ one can find a $k'$ and $m'$ so that $\equiv^R_{k',m',\Omega}$ is a refinement of $\approx^k_{m,\Omega}$? One way this might be established is by showing that the relations $\approx^k_{m,\Omega}$ are themselves definable in IFPR. If it turns out that this is not the case, then we would have established that there is a PTIME property not in in IFPR. A natural line of investigation would then be to extract from the invertible-map game a suitable logical operator, stronger than the matrix-rank operator, that is characterised by this game.

A more general direction of research that is suggested by this work is to explore other partition games which can be defined by suitable equivalence conditions on the partition matrices. There is space here for defining new logics and also new isomorphism tests.

**Acknowledgements.** This research was partially done whilst the authors were participants in the programme *Semantics and Syntax* at the Isaac Newton Institute for the Mathematical Sciences in Cambridge, UK.